\documentclass[11pt,tightenlines,eqsecnum,floats,aps,amsmath,amssymb,nofootinbib,prd,floatfix]{revtex4}

\usepackage{graphicx}

\usepackage{array}
\usepackage{amsmath}
\usepackage{amsfonts}
\usepackage{amssymb}
\usepackage{empheq}
\usepackage{color}
\usepackage{colortbl}
\usepackage{mathrsfs}

\definecolor{LightGreen}{rgb}{0, 0.8, 0.2}
\definecolor{LightRed}{rgb}{0.8, 0, 0}
\definecolor{LightBlue}{rgb}{0.0, 0, 0.8}

\def\be{\begin{equation}}
\def\ee{\end{equation}}
\def\ba{\begin{eqnarray}}
\def\ea{\end{eqnarray}}

\newcommand{\pb}[1]{\hbox{\lower0.5ex\hbox{${}_{\leftarrow}$}}\kern-1.9ex{#1}}

\def\={\,\hat{=}\,}
\def\out{{\rm out}}
\def\inn{{\rm in}}

\def\TF{{\rm TF}}
\def\g{\tau} 

\def\ut#1{\rlap{\lower1ex\hbox{$\sim$}}{#1}}

\def\f{\frac}
\def\rmd{{\textrm{d}}}

\def\={\hat{=}}

\def\t{\tilde}

\def\ub{\underbar}
\def\ul{\underline}
\def\h{\hat}
\def\b{\bar}
\def\vk{\vec k}
\def\vp{\vec p}

\def\Do{\mathring{D}}

\def\scri{\mathcal{I}}
\def\scrip{\mathcal{I}^{+}}
\def\scrim{\mathcal{I}^{-}}
\def\L{\mathcal{L}}
\def\S{\mathcal{S}}
\def\T{\mathcal{T}}

\def\F{\mathcal{F}}
\def\H{\mathcal{H}}
\def\Hs{{\underline{\mathcal{H}}}}
\def\P{\mathcal{P}}
\def\A{\mathcal{A}}

\usepackage{graphicx,verbatim}
\usepackage{amsmath}
\usepackage{amsfonts}
\usepackage{amssymb}
\usepackage{color}

\def\rmd{{\mathrm{d}}}

\def\B{\mathfrak{B}}

\def\V{\mathbf{V}}

\def\Gammabf{\mathbf{\Gamma}}
\def\Omegabf{\mathbf{\Omega}}
\def\Sbb{\mathbb{S}}
\def\Rbb{\mathbb{R}}

\def\starK{{}^{\star}\!K}

\def\Do{\mathring{D}}

\def\SO(3){\rm SO(3)}
\def\so(3){\rm so(3)}
\def\SO(4){\rm SO(4)}
\def\so(4){\rm so(4)}
\def\SO(1,4){\rm SO(1,4)}
\def\so(1,4){\rm so(1,4)}
\def\SU(2){\rm SU(2)}

\def\be{\begin{equation}}
\def\ee{\end{equation}}
\def\ba{\begin{eqnarray}} 
\def\ea{\end{eqnarray}}

\begin{document}

\title{Null infinity, the BMS group and infrared issues}
\author{Abhay Ashtekar}
\email{ashtekar.gravity@gmail.com} \affiliation{Institute for
Gravitation and the Cosmos \& Physics
  Department, Penn State, University Park, PA 16802, U.S.A.}
\author{Miguel Campiglia}
\email{campi@fisica.edu.uy  } \affiliation{ Instituto de F\'isica, Facultad de Ciencias, Igu\'a 4225, 
              esq. Mataojo, 11400 Montevideo, Uruguay.}
\author{Alok Laddha}
\email{laddha@cmi.ac.in} \affiliation{ Chennai Mathematical Institute, Siruseri, Chennai, India}

\begin{abstract}

There has been a recent resurgence of interest in the structure of the gravitational field at null infinity, sparked by new results on soft charges and infrared issues related to the S matrix theory in perturbative quantum gravity. We summarize these developments and put them in the broader context of research in the relativity community that dates back to several decades. In keeping with intent of this series, this overview is addressed to gravitational scientists who are not experts in this specific area.

\end{abstract}
\maketitle
\section{Introduction}
\label{s1}

In the mid 1960s Penrose \cite{rp} introduced conformal completions of asymptotically flat space-times, thereby `geometrizing'  the Bondi-Sachs description of gravitational waves \cite{bondi,bondi-sachs}. In particular, the null boundary $\scri$ of the physical space-time provided by this framework serves as the natural arena for describing radiative aspects of  zero rest mass fields. In recent years, numerical simulations of binary coalescences have routinely used the ensuing gauge invariant framework to describe gravitational waves in full, non-linear general relativity. As we will show in this article, the $\scri$-framework also provides a useful platform  to analyze subtle and unforeseen mathematical issues  associated with gravitational radiation both in classical and quantum gravity. 

Specifically, we will explain the following features in general terms, emphasizing  structures that lie at the heart of the recent resurgence of interest: \\
(i) An invariant characterization of gravitational waves in terms of natural connections and their curvature on $\scri$. This description brings radiative aspects of general relativity closer to those of non-Abelian gauge theories;\\
 (ii) Enlargement of the Poincar\'e group ${\P}$ to the infinite dimensional Bondi-Metzner Sachs (BMS) group $\B$ in space-times that admit gravitational radiation, and the natural reduction of $\B$ to $\P$ in absence of radiation; \\
 (iii) Presence of an infinite family $\V$ of distinct connections on $\scri$ with trivial curvature.  In the terminology used in non-Abelian gauge theories, these are \emph{`non-trivial classical vacua'};\\
 (iv) A natural correspondence between the space $\mathbf{P}$ of Poincar\'e subgroups of $\B$ and the space $\V$ of non-trivial vacua;\\
{(v)} Emergence of the memory effect as a direct consequence of the classical vacuum degeneracy;\\
{(vi)} Encoding of the two  radiative modes of the full, non-linear gravitational field in connections at $\scri$, and emergence of formulas for fluxes of energy momentum, supermomentum and angular momentum as Hamiltonians corresponding to the generators of the BMS group $\B$ on the phase space of these radiative modes;\\
(vii) Quantization of radiative modes of full non-linear general relativity and precise association of mass and spin to these quanta;\\
{(viii)} Relation between non-trivial classical vacua and emergence of infrared sectors in the quantum theory. These are essential to make the perturbative $S$-matrix well-defined; and \\
{(ix)} Correspondence between  Weinberg's soft graviton theorems and new conservation laws in perturbative quantum gravity. \smallskip
 
Literature on this subject is vast and spans several decades. We will only be able to sketch the basic ideas; details can be found in review articles such as \cite{aa-bib,aa-yau,as-book} and 
references therein. (Because of space limitation, our list of references to original papers is necessarily incomplete.) Our goal is only to provide a reasonably self-contained summary of the status of these ideas.  In view of the intended audience, the material will be presented from the viewpoint of relativists rather than particle physicists, with emphasis on geometry and methods used by relativists. 
 
 \section{The BMS Group $\B$}
 \label{s2}
 
 To study isolated systems emitting gravitational waves, Bondi, Sachs and others restricted themselves to asymptotically flat space-times. Specifically, they assumed that the physical metric $g_{ab}$  approaches a Minkowski metric $\eta_{ab}$ as $1/r$ if one recedes from sources in null directions  (where $r$ is a radial coordinate of $\eta_{ab}$).  With these boundary conditions, one would have expected the asymptotic symmetry group to be just the Poincar\'e group $\P$. Indeed, this was routinely assumed in particle physics literature (e.g. in perturbative quantum gravity, first presented by Feynman in the same conference that Bondi and others first presented their results \cite{bondi}). It was a major surprise that this was \emph{not the case:} Rather, the asymptotic symmetry group $\B$ is an infinite dimensional generalization of $\P$.
 
Intuitively, one can trace back the emergence of this enlargement to the following fact.  Consider  a diffeomorphism $ t \to t^{\prime} = t +f(\theta,\phi), \,\, \vec{x} \to \vec{x}^{\prime} = \vec{x}$ where $t,\vec{x}$ are Cartesian coordinates of $\eta_{ab}$. This is an \emph{angle dependent  translation}, whence the metric  $\eta_{ab}$ is sent to a \emph{distinct} flat metric $\eta^{\prime}_{ab}$. A detailed examination showed that if a physical metric $g_{ab}$ approaches $\eta_{ab}$ as $1/r$ in the manner specified by Bondi et al, then it also approaches $\eta^{\prime}_{ab}$ as $1/r^{\prime}$. But since the two Minkowski metrics are distinct, their isometry groups $\P$ and $\P'$ are also distinct. Therefore the asymptotic symmetry group associated with the physical  metrics $g_{ab}$ under considerations must include \emph{both} these Poincar\'e groups. The argument continues to hold if the `angle dependent translation' we considered is space-like or null. The BMS group $\B$ can be interpreted as the `union' of Poincar\'e groups associated with all these Minkowski metrics.

To make these considerations precise, let us begin with a mathematical characterization of space-times under consideration  \cite{rg,aabs,aa-yau}.\\
\textbf{Definition 1:} A space-time $(\hat{M},\hat{g}_{ab})$ is said to be \textit{asymptotically Minkowskian} at null infinity if there exists a manifold $M$ with boundary $\scri$ equipped with a metric $g_{ab}$, and a diffeomorphism from $\hat{M}$ onto  the interior $M\, \setminus\, \scri$ (with which we identify $\hat{M}$ and $M\, \setminus\, \scri$) such that:%
\footnote{Throughout our discussion $\scri$ will stand for either the future null infinity $\scrip$, or the past, $\scrim$. However, in classical considerations  primary interest lies in $\scrip$.}
\\
(i)\, there exists a smooth function $\Omega$ on $M$ with $g_{ab}=\Omega^2 \hat{g}_{ab}$ on $\hat{M}$; $\Omega=0$ on $\scri$ and $n_a := \nabla_a \Omega$ is nowhere vanishing on $\scri$;\\
(ii) $\scri$ is topologically $\Sbb^2\times \Rbb$;\\
(iii)\,$\hat{g}_{ab}$ satisfies Einstein's equations $\hat{R}_{ab} - \frac{1}{2} \hat{R} \hat{g}_{ab}  = 8 \pi G \; \hat{T}_{ab}$, where $\Omega^{-2} \hat{T}_{ab}$ has a smooth limit to $\scri$; and,\\
(iv) The integral curves of $n^{a}$ are complete on $\scri$ for any choice of the conformal factor for which makes $\scri$ divergence-free (i.e. $\nabla_{a}n^{a} = 0$ on $\scri$).

These conditions can be motivated as follows. Condition  (i) ensures that the conformal factor $\Omega$ falls off as $1/r$ as one recedes from sources in null directions;\, (ii) ensures that we encompass all angular directions in the process;\, ( iii)  arises from the known asymptotic behavior of physically interesting zero rest mass fields in Minkowski space; and the global (in time) condition  (iv)  ensure that the action of the BMS group $\B$ is well-defined on $\scri$, rather than just its Lie algebra. Note that contrary to the first treatments \cite{rp}, the definition does \emph{not} require null geodesics in the physical space-time to have end points on $\scri$. (In particular, space-times can have black holes). Nonetheless the theory of gravitational waves can  be fully developed.

 A large class of physically interesting stationary space-times  representing isolated bodies satisfy these conditions. For space-times admitting gravitational radiation, there are \emph{global} existence results for small \emph{non-linear} fluctuations around Minkowski space-time (see, in particular, \cite{dcsk,hf1,pced,lb,nz,hf2}). However,  in full general relativity so far there are no analytic results for large data that, e.g., lead to gravitational collapse, nor for solutions with compact bodies as sources (for which the theory was first developed!). There do exist a large class of numerical simulations that strongly indicate that the conditions are satisfied in solutions with sources, e.g., in binary coalescences. 
 
Using this definition, one can provide a precise characterization of the BMS group.  For simplicity, throughout this article, we will follow the mainstream literature and \emph{restrict ourselves to divergence-free conformal frames.} Then, every space-time satisfying this definition has the following structure \cite{rp,aa-yau}):\\ 
(1) The conformal boundary $\scri$ is a null 3-manifold;\\ 
(2)  We have a pairs of fields $(q_{ab},  n^{a})$, defined intrinsically on $\scri$, where $q_{ab}$ is the degenerate  metric with signature 0,+,+ and $n^{a}= g^{ab}\nabla_{b}\Omega$ the null normal to $\scri$. They satisfy $\mathcal{L}_{n} q_{ab} =0$ and $q_{ab} n^{b} =0$;\\
(3) Permissible conformal rescalings send $(q_{ab}, n^{a}) \to (\omega^{2}q_{ab}, \omega^{-1} n^{a})$, where $\mathcal{L}_{n} \omega =0$. 

Because a 2-sphere admits a unique conformal structure, asymptotically Minkowskian space-times admit a universal structure: the manifold $\scri$, equipped with equivalence classes $\{(q_{ab}, n^{a})\}$ of pairs $(q_{ab}, n^{a})$,  where two pairs are equivalent if they are related by conformal rescalings given above under (3). \emph{The asymptotic symmetry group of these space-times is then the subgroup of the diffeomorphism group ${\rm Diff} (\scri)$ of $\scri$ that preserves this universal structure.}  This is the BMS group $\B$. To spell out its structure, let us first recall that Poincar\'e group $\P$ is a semi-direct product, $P = \L \ltimes \T$: the (Abelian) translation group $\T$ is a normal subgroup of $\P$ and the quotient $\P/\T$ is the Lorentz group $\L$. Similarly, $\B$ is a semi-direct product {$\B = \L  \ltimes \S$}: the 4-dimensional translation group $\T$ is replaced by the infinite dimensional (Abelian) group $\S$ of supertranslations, generated by vector fields $fn^{a}$ on $\scri$ where $f$ is a scalar satisfying $\mathcal{L}_{n} f=0$. (Since $n^{a} \to \omega^{-1} n^{a}$ under permissible conformal rescalings, $f \to \omega f$; so $f$ has conformal weight +1. This important point is sometimes overlooked.) 

Interestingly, $\B$ also admits a \emph{unique} 4-dimensional Abelian normal subgroup $\T \subset \S$ which coincides with the group of translations if the underlying physical space-time is Minkowski space \cite{sachs}. This fact is extremely useful in practice: in the classical theory, it enables us to define fluxes of energy momentum across $\scri$ as Hamiltonians generating translations, and in the quantum theory it enables us to unambiguously decompose fields into creation and annihilation operators. (In a so-called \emph{Bondi conformal frame} --in which $q_{ab}$ is the unit 2-sphere metric-- $\T$ is generated by vector fields  $f n^{a}$ where $f= c_{o} Y_{00} + c_{m} Y_{1,m}$ is a linear combination of the first 4 spherical harmonics.) \medskip

\emph{Remark:} For simplicity we assume that all fields under consideration are smooth everywhere, including $\scri$. This, in particular,  implies that the asymptotic curvature tensor `peels' in the standard manner \cite{rp,etnrp1}. It is sometimes argued that this assumption is too strong because it is not satisfied, e.g.,  in the Christodoulou-Klainnerman analysis \cite{dcsk}. However, with a suitable choice of the class of initial data, and one can achieve a $C^{k}$ differentiability at $\scri$ for any $k$ (see, e.g. \cite{pced}). Furthermore since all these global existence results have been established only for \emph{source-free solutions} with `small data', a priori it is difficult to decide which of these sets of initial data is more appropriate in physical situations, e.g., for binary coalescences and scattering.
 
 \section{Radiative modes of full, non-linear general relativity}
 \label{s3}
 Null infinity provides a natural arena for analyzing properties of radiation of zero rest mass fields in the classical theory and for constructing the Hilbert space of asymptotic states for the quantum S-matrix theory. For example, for gauge fields on \emph{curved}, asymptotically flat space-times  one cannot carry out Fourier transforms, but the $\scri$-arena enables us to extract the radiative modes in a gauge invariant manner in the classical theory, and decompose the field operators  into creation and annihilation parts in the quantum theory (see, e.g.,  \cite{aa-bib,as-book}). In this section we will see that the situation is the same for full non-linear gravity. 
\subsection{Gravitational connections on $\scri$}
 \label{s3.1}
Consider an asymptotically Minkowskian space-time $(\hat{M}, \hat{g}_{ab})$ and a conformal completion $(M,g_{ab})$ thereof. Since the intrinsic metric $q_{ab}$ on $\scri$ is degenerate, $\scri$ admits an infinite number of intrinsic connections $D$ that are metric compatible, i.e., satisfy $D_{a} q_{bc} =0$. However, as a consequence of our boundary conditions, the null normal $n^{a}$ to $\scri$ has zero twist, expansion, and shear in $(M,g_{ab})$. As a consequence the connection $\nabla$  on $M$ (satisfying $\nabla_{a}g_{bc} =0$) induces a unique intrinsic connection $D$ on $\scri$  via pull-back. It satisfies 
\be \label{1} D_{a} q_{bc} =0, \quad {\rm and} \quad D_{a} n^{b} =0. \ee
Recall that $\{(q_{ab}, n^{a})\}$ are `universal' --they refer to the entire class of asymptotically flat space-times. The connection $D$, on the other hand, does encode interesting information about the specific space-time under consideration. As we now discuss, this information turns out to be sufficient to characterize \emph{precisely} the radiative content of space-time \cite{aa-rad}.

Let us begin by considering conformal rescalings $g_{ab} \to g^{\prime}_{ab} = \omega^{2} g_{ab}$ where $\omega =1$ \emph{at} $\scri$. Since  $g_{ab}|_{\scri}$ does not change under this rescaling, neither does the pair $(q_{ab}, n^{a})$. However, $\nabla$ --and hence its pull-back $D$ to $\scri$-- can change if $\nabla_{\!a} \omega \not= 0$ at $\scri$. Since $g_{ab}$ and $g^{\prime}_{ab}$ both refer to the same physical metric $\hat{g}_{ab}$ these rescalings are `gauge transformations' in the sense that they can have no physical effect. Therefore we are led to consider equivalence classes $\{D\}$ of connections related to each other by these rescalings:
It turns out that we need to regard $D$ and $ D^{\prime}$  as equivalent if and only if 
\be \label{2}
 (D^{\prime}_{a} -  D_{a}) k_{b} =  (n^{c}k_{c})\, q_{ab}\quad \hbox{\rm{for all 1-forms} $k_{a}$ on $\scri$.} \ee
Let us now consider curvature of these connections \cite{aa-rad,aa-bib,aa-yau}. Since we are on a 3-manifold, the Riemann tensor  {is} determined completely by a second rank tensor $S_{a}{}^{b}$, which, however, changes under remaining conformal freedom.  The conformally invariant part  is encoded in two symmetric second rank tensor fields on $\scri$. The first is the Bondi news tensor $N_{ab}$ which is transverse and traceless, i.e., satisfies  $N_{ab} n^{a}=0$ and $N_{ab} q^{ab} =0$.%
\footnote{Since $q_{ab}$ is degenerate, $q^{ab}$ is defined only up to addition of terms of the form $t^{(a} n^{b)}$ where $t^{a}$ is tangential to $\scri$. However, this ambiguity does not matter in the traceless condition because $N_{ab}n^{b}=0$.}
If one goes to a \emph{Bondi conformal frame}, then $N_{ab}$ is simply the trace-free part of $S_{ab} := S_{a}{}^{c} q_{bc}$. The second tensor $\starK^{ab}$ is given by
\be  \label{3}
 D_{[a} S_{b]}{}^{c} = \f{1}{4}\,  \epsilon_{abm}  \starK^{mc} \ee
 where $\epsilon_{abm}$ is the alternating tensor (i.e., the volume 3 form) on $\scri$ induced by $g_{ab}$.  The field $\starK^{ab}$ is trace-free and captures 5 of the 10 components of the asymptotic Weyl curvature of $g_{ab}$. In the Newman Penrose language \cite{etnrp1}, they are given by $\Psi_{4}^{o}, \, \Psi_{3}^{o},\, {\rm Im}\,\Psi_{2}^{o}$ that are associated with the radiative aspect of the gravitational field \cite{rp,aa-rad,aa-bib}. The `Coulombic part' of the asymptotic curvature  --$\Psi_{2}^{o}$ that features in the expression of the Bondi 4-momentum \cite{bondi-sachs,rp} $\Psi_{1}^{o}$ that features in the expression of the angular momentum \cite{tdms,ak}-- are not encoded in the gravitational connection $D$. 
 
 Indeed, one can show that $\{D\}$ only captures the two transverse-traceless modes of the asymptotic gravitational field as follows. Fix a pair $(q_{ab}, n^{a})$ at $\scri$ and consider conformal completions $(M,g_{ab})$ of various space-times that induce this pair on their conformal boundary. Each of these space-times provides a connection $D$ on $\scri$. How many distinct $D$ can we have? Using the fact that each of these connections satisfies $D_{a}q_{bc}=0$ and $D_{a} n^{b} =0$,  it is easy to show that the difference between any two is of the type: $(\tilde{D}_{a}- D_{a}) k_{b} = (n^{c}k_{c}) \,\Sigma_{ab} $ where $\Sigma_{ab}$ is a symmetric tensor field, transverse to $n^{a}$. Finally, the conformal freedom of Eq. (\ref{2}) implies that the trace-part of $\Sigma_{ab}$ is pure gauge, whence two gauge \emph{inequivalent} $\{\tilde{D}\}$ and $\{D\}$ are related by trace-free part $\sigma_{ab}$ of $\Sigma_{ab}$. The fields $\sigma_{ab}$ represent the \emph{two transverse traceless radiative modes} of the gravitational field in \emph{full, nonlinear general relativity} \cite{aa-rad,aa-bib}.
 
Let us summarize. The pairs $(q_{ab}, n^{a})$ on $\scri$ are `universal', common to all space-times under consideration. By contrast, different  space-times induce different (equivalence classes) of connections $\{D\}$ on $\scri$.  The space $\Gammabf$ of all $\{D\}$'s has the structure of an affine space (inherited from that of all connections $D$). If we choose any one $\{D\}$ as the origin, then $\Gammabf$ can be coordinatized by transverse traceless tensor fields $\sigma_{ab}$. As we will see  in section \ref{s3.2}, they can be interpreted as `shear tensors' if the origin in $\Gammabf$ is chosen suitably.
 
\subsection{Classical vacua and supertranslations}
\label{s3.2}

In Yang-Mills theories, connections with trivial curvature are called `classical vacua'.  Let us use the same terminology here. For gravitational connections $\{D\}$, curvature is encoded in $N_{ab}$ and $\starK^{ab}$. As we will see, it is the News tensor $N_{ab}$ that directly governs fluxes of BMS momenta --energy momentum, supermomentum and angular momentum-- carried away by gravitational waves. However, one can show that vanishing of $\starK^{ab}$ implies  vanishing of $N_{ab}$ \cite{abk1} but not vice versa. Therefore, \emph{a  connection $\{\Do\}$ is said to represent a classical gravitational vacuum if the corresponding ${\starK}^{ab}$ vanishes.} If the connection induced by a space-time  geometry on $\scri$ is a vacuum $\{\Do\}$ on $\scrip$ (respectively, $\scrim$), then the flux of \emph{all} BMS momenta across $\scrip$ (respectively, $\scrim$) vanishes. Indeed, the simplest way to specify the `no incoming radiation' condition for isolated gravitating systems is to require vanishing of $\starK^{ab}$ at $\scrim$. In this respect, the situation is completely parallel to gauge theories.

However, there is also an obvious difference between gauge theories and gravity: Distinct vacua $\{\Do\}$ are now related to each other by  asymptotic \emph{space-time} symmetries, rather than by asymptotic internal gauge transformations. One can show that each vacuum $\{\Do\}$ is left invariant by the action of the 4-dimensional translational sub-group $\T$ of $\B$, but not by a supertranslation (that is not a translation). Thus, there are `as many' classical vacua $\{ \Do \}$ as there are `pure' supertranslations. In fact the quotient $\S/\T$ acts {on} the space $\V$ simply and transitively. Since $\S/\T$ also acts simply and transitively on the space $\mathbf{P}$ of Poincar\'e subgroups  $\P$ of the BMS group $\B$, one would expect a close relation between $\V$ and $\mathbf{P}$. And indeed, there is one: each $\{\Do\} \in \V$ is left invariant precisely by one Poincar\'e subgroup $\P \in \mathbf{P}$ \cite{aa-rad}.  In this precise sense, \emph{the enlargement of the Poincar\'e group to the BMS group can be traced back to  vacuum degeneracy of gravitational connections.}

There is a convenient characterization of these vacua.  Consider a conformal frame $(q_{ab}, n^{a})$, denote by $u$ an affine parameter of $n^{a}$ (so $\mathcal{L}_{n} u =1$) and set $\ell_{a} = -D_{a}u$.  For each such $\ell_{a}$, one can show that there is a unique classical vacuum $\{\Do\}$ on $\scri$ such that $\Do_{a}\, \ell_{b} \propto q_{ab}$ for all $\Do \in \{\Do\}$.   Thus, the $u={\rm const}$ cross-sections of $\scri$ \emph{are all shear-free with respect to the classical vacuum}  $\{\Do\}$:\,\, $\mathring\sigma_{ab} := \Do_{a} \ell_{b}  - \f{1}{2} q^{cd} (\Do_{c}\ell_{d})\, q_{ab} =0$ for all $\Do \in \{\Do\}$. Next, let us consider BMS \emph{translations}, i.e. elements of  $\T$. Recall that each classical vacuum is left invariant by translations and that, in a Bondi conformal frame, translations  correspond to diffeomorphisms $u\to \t{u} = u + c_{o} Y_{0,0} + c_{m} Y_{1,m} (\theta,\varphi)$ on $\scri$. One can show that the 4-parameter family of cross sections $\t{u} = {\rm const}$ are also shear-free with respect to the $\{\Do\}$ we began with.  Thus, there is a 1-1 correspondence between classical vacua $\{\Do\}$ and 4-parameter families of cross-sections that are shear-free with respect to them \cite{etnrp2,aa-rad,aa-bib}.

Note, however, that the freedom in the choice of the affine parameter $u$ is much larger, given 
by $u \to u + f$ where $\mathcal{L}_{n} f =0$. This is precisely the supertranslation freedom. Thus, under the action of a supertranslation $u \to u^{\prime}=u + f$, we have  $\ell^{a} \to \ell^{\prime}_{a}$, and $\ell^{\prime}_{a}$ defines a new classical vacuum $\{\Do^{\prime}\}$ with respect to which the cross-sections $u^{\prime} = {\rm const}$ are shear-free. As one would expect,  $\{\Do^{\prime}\}$  is precisely the image of  $\{\Do \}$ under the given supertranslation. 

To summarize,  each classical vacuum $\{\Do\} \in \V$ singles out a 4-parameter family of cross-sections of $\scri$ that are shear-free with respect to it.  If there is no gravitational radiation across $\scri$ \, --i.e. if we are given a space-time which induces  a classical vacuum $\{\Do\}$ on $\scri$--\,  then we also have a 4-parameter family of shear-free cross-sections. The BMS transformations that leave this family invariant constitute a Poincar\'e subgroup $\P$ of $\B$ --precisely the one that leaves the given $\{\Do\}$ invariant.  In this precise sense, the \emph{enlargement of the Poincar\'e group $\P$ to the BMS group $\B$ can be traced directly to the presence of gravitational waves.} In the more general case when there is radiation at $\scri$,  the induced connection $\{D\}$ at $\scri$ has non-trivial curvature. However one demands that the curvature falls-off as $u \to \pm \infty$ --i.e., as we approach $i^{+}$ and $i^{o}$ along $\scrip$ (or  $i^{o}$ and $i^{-}$ along $\scrim$).  Thus every connection $\{D\}\in\Gammabf$ is assumed to approache classical vacua $\{{\Do}^{\pm} \}$ at the two ends. However, generically the two vacua are \emph{distinct}. We will see that this is the origin of the `memory effect' in the classical theory and subtle infrared issues in the quantum theory.

\emph{Remark:}  Following the procedure used in the literature, we used the condition $\starK^{ab} =0$ to characterize `trivial curvature'. It implies that the News tensor $N_{ab}$ must vanish.  But the converse is not true. Now, as we will see in section \ref{s3.3}, the weaker condition  $N_{ab}=0$ suffices to guarantee that fluxes of \emph{all} BMS momenta vanish across $\scri$. Therefore one could envisage enlarging the notion of classical vacua by asking only $N_{ab}$ to vanish. In the Newman-Penrose notation this would correspond to requiring only $\Psi_{4}^{o} =0$ and $\Psi_{3}^{o} =0$ on $\scri$\, --the component ${\rm Im} \,\Psi_{2}^{o}$ need not vanish. This leads to a more general framework, allowing for a `magnetic type memory' in the classical theory and a new `magnetic family' of infrared sectors in the quantum theory. However, it is not yet clear if whether this generalization has a physical basis \cite{jw}.  
   
\subsection{Phase space of radiative modes}
\label{s3.3}

In physical theories on Minkowski space-time one generally obtains expressions of energy-momentum and angular momentum using the stress-energy tensor and space-time Killing fields. However, the same expressions can be obtained as Hamiltonians generating canonical transformations induced by the action of the Poincar\'e group. In general relativity, there is no stress-energy tensor associated with the gravitational field itself nor, in general, Killing vectors. 
But one nonetheless still use a phase space framework in conjunction with asymptotic symmetries to obtain Hamiltonians to unravel the physical content of various solutions. For example, if one works with space-times that are asymptotically flat at spatial infinity, Hamiltonians generating asymptotic translations provide us with the expression of the Arnowitt-Deser-Misner energy-momentum. We will now show that the situation is similar at null infinity:
One can construct a phase space of radiative modes and the expressions of Hamiltonians generating BMS symmetries provide us with expressions of  fluxes of the BMS momenta across $\scri$. (For further details, see \cite{aaam,aams,aa-bib}.)

Recall first that starting from the Lagrangian of any physical system, there is a systematic procedure to introduce a symplectic structure on the space of solutions to the theory. This is the covariant phase space, that does not require slicing the space-time into space and time (see, e.g., \cite{abr}). Let us then consider the covariant phase space $\Gammabf_{\rm cov}$ of general relativity, consisting of solutions that are asymptotically Minkowskian, and which  admit (asymptotically flat) Cauchy surfaces $M$ with topology $\mathbb{R}^{3}$ (without internal boundaries).  $\Gammabf_{\rm cov}$ is endowed with a natural symplectic structure $\Omegabf$. Since it is a 2-form on $\Gammabf_{\rm cov}$, given any given a vacuum solution $\h{g}_{ab}$ and two linearized solutions $\delta\h{g}_{ab}, \t{\delta}\h{g}_{ab}$ thereon, it yields a number $\Omegabf\mid_{\h{g}}\, (\delta, \t\delta)$ {which can be} expressed as a (conserved) integral on any Cauchy surface $M$.  Now, each $\h{g}_{ab} \in \Gammabf_{\rm cov}$ induces a radiative mode $\{D\}$ on $\scri$, and each linearized solution $\delta \h{g}_{ab}$, a linearized connection $\delta\{D\}$, naturally represented by a symmetric, transverse tensor field $\gamma_{ab}$ on $\scri$. (In terms of concepts introduced in section \ref{s3.2}, $\gamma_{ab}$ has the interpretation of `linearized shear'.) By taking the limit as the Cauchy surface $M$ approaches $\scri$,  and astutely using the gauge freedom \cite{rgbx}, one can cast the expression of the symplectic structure in terms of structures available at $\scri$\, \cite{aaam}:
\ba \lim_{M\to \scri} \Omegabf\mid_{\h{g}}\, (\delta, \t\delta) &=& \f{1}{8\pi G}\,\int_{\scri} {\rm d}^{3}\scri\,\big(\gamma_{ab} \L_{n} \t\gamma_{cd} - \t\gamma_{ab} \L_{n} \gamma_{cd}\big)\, q^{ac} q^{ad}\,  \nonumber\\
&=:& \Omegabf\mid_{\{D\}}\, (\gamma, \t\gamma)  \label{4} \ea
These considerations motivate the introduction of the phase space of radiative modes. To ensure that various integrals on $\scri$ converge, one restricts oneself to connections $\{D\}$ whose curvature tensors $N_{ab}, \starK^{ab}$ fall off as $1/|u|^{(1+\epsilon)}$ as $u\to \pm \infty$ along $\scri$, for some $\epsilon >0$. Then the shear tensors $\sigma_{ab}^{\pm}$ that characterize  the rate at which $\{D\}$ approaches vacuum configurations $\{\Do^{\pm}\}$ at the two ends of $\scri$ remain bounded as $u \to \pm \infty$. In particular, these conditions imply that fields $\gamma_{ab}$ in Eq. (\ref{4}) remain bounded and $(\L_{n}\gamma_{ab})$ falls off as $1/|u|^{(1+\epsilon)}$ as $u\to \pm \infty$.  The space $\Gammabf$ of these connections, equipped with the  symplectic structure $\Omegabf$ of Eq. (\ref{4}) is the \emph{phase space of radiative modes of full non-linear general relativity.} (For details on the radiative phase space, see \cite{aams}.)

Since the BMS group preserves the universal structure at $\scri$, it has a natural action on $\Gammabf$. One can show that it preserves the symplectic structure.  Consider a generic generator $\xi^{a}$ of the BMS group: $\L_{\xi} q_{ab} = 2\alpha q_{ab}$ and $\L_{\xi} n^{a} = -\alpha n^{a}$ with $\L_{n} \alpha =0$. The Hamiltonian $H_{\xi}$ generating the corresponding symplectomorphism (i.e. canonical transformation) on $\Gammabf$  is given by
\be H_{\xi} = \f{1}{16\pi G} \int_{\scri}  {\rm d}^{3}\scri \, N_{ab} \big[(\L_{\xi} D_{c} - D_{c} \L_{\xi})\ell_{d} + 2 N_{ab} \ell_{c} D_{d} \alpha \big]\, q^{ac} q^{bd}\,  , \label{5}\ee
where $\ell_{a}$ is any 1-form on $\scri$ satisfying $\ell_{a} n^{a} = -1$. $H_{\xi}$ has has the interpretation of the \emph{total flux across $\scri$ of the BMS momentum corresponding to $\xi^{a}$.} For our purposes, it will suffice to restrict $\xi^{a}$ to be BMS supertranslations: $\xi^{a} = f n^{a}$ where $\L_{n} f =0$. (Recall that $f$ has conformal weight 1, i.e., under $(q_{ab}, n^{a})\, \to \, (\omega^{2}q_{ab}, \omega^{-1}n^{a})$, we have $f \to \omega f$.) Then,  $\alpha=0$ and the corresponding Hamiltonian $H_{f}$ simplifies to :
\ba  \label{6} H_{f} &=& \f{1}{16\pi G} \int_{\scri} {\rm{d}}^{3}\scri \,
N_{ab} \big(f\, S_{cd} + D_{c} D_{d} f \big) q^{ac} q^{bd}\,  \nonumber\\
&=& \f{1}{16\pi G} \int_{\scri} {\rm{d}}^{3}\scri\,  N_{ab} \big(f\, N_{cd} + D_{c} D_{c} f \big) q^{ac} q^{bd}\,  \quad \hbox{\rm in a Bondi conformal frame,}  \nonumber\\
&=& Q_{f} ({\rm hard})+ Q_{f} (\rm soft)\, , \ea
where in the second step we have used the fact that the news tensor $N_{ab}$ is the trace-free part of $S_{ab}$ in a Bondi conformal frame, and in the third, denoted the terms quadratic and linear in News as `hard' and `soft' charges, following recent terminology \cite{as1}. If $\xi^{a}$ is a BMS \emph{translation}, then $D_{a}D_{b}f$ is proportional to the metric and the `soft' part vanishes. If furthermore  $\xi^{a}$ is a {time}  translation, then $f$ is positive, whence the flux of energy carried by gravitational waves is \emph{manifestly positive}. Historically,  there was considerable  controversy on the reality of gravitational waves in full, non-linear general relativity \cite{dk} and this positivity played an important role in resolving this issue. 
 
 \emph{Remarks:}\\
1. The derivation \cite{aaam} of (\ref{4}) predates the global stability analyses \cite{dcsk,hf1,pced} and did not take into account functional analytic issues; it simply assumed the existence of vacuum solutions with physically motivated properties. It would be of interest to put it on a more rigorous footing.\\ 
 2. Note that $\Gammabf$ is an affine space. It is tempting to choose a vacuum configuration $\{\Do\}$ as the origin and endow a vector space structure to simplify calculations of fluxes. However, since no vacuum is left invariant under `pure' supertranslations, this strategy is not viable and indeed had led to incorrect expressions of supermomentum fluxes in the early literature, where the second term in Eq. (\ref{6}), $Q_{f}(\rm soft)$,  was missing .  We will find that this term plays a key role in the discussion of infrared issues.\\
 3. The covariant phase space $\Gammabf_{\rm cov}$ we began with was arrived at by using source-free solutions and the underlying assumptions also excluded black holes. The expression of fluxes of BMS momenta carried by gravitational waves across $\scri$ was derived using this phase space \cite{aams}.  Nonetheless, it is generally assumed that these expressions are valid \emph{also in presence of sources and black holes.} (This is analogous to the fact that  although the expression of the ADM 4-momentum at spatial infinity was initially derived using the Hamiltonian formulation of vacuum general relativity and in absence of internal boundaries representing black holes, it is interpreted as the total 4-momentum of the system even when these assumptions are violated.)  However, some recent results \cite{aabb} suggest that this assumption may have to  be reexamined for angular momentum fluxes across $\scri$. \\
4.  In presence of massive particles in space-time, there is an additional contribution to the hard charge $Q_{f}(\rm hard)$ that is not captured in the  gravitational waves traversing $\scri$. Unfortunately, due to space limitation we will not be able to discuss this issue.
 \section{Quantum theory: The Fock representation} 
\label{s4}
In the classical theory, radiative modes $\{D\}$ constitute  physically admissible states --i.e., points of the radiative phase space $\Gammabf$-- if the corresponding curvature tensors $N_{ab}, \starK^{ab}$ fall off as $1/|u|^{1+\epsilon}$ as $u \to \pm \infty$. This condition ensures, in particular,  that the total fluxes of BMS momenta $H_{\xi}$ across $\scri$ are finite. Recall that the connection $\{D\}$ can tend to two \emph{distinct} vacua $\{\Do^{\pm}\}$ at the two ends of  $\scri$.  Given such a $\{D\}$, consider any family of cross-sections of $\scri$, related to each other by a BMS time translation and calculate the shear $\sigma_{ab}$ of each of these cross-sections.  Then the relation between  the two vacua $\{\Do^{\pm}\}$ can be expressed in terms asymptotic shears of these cross sections as follows: If $u \to u + f(\theta,\varphi)$ is the supertranslation that maps $\{\Do^{-}\}$ to $\{\Do^{+}\}$, then $[\sigma] := \big(\sigma_{ab}|_{ u=\infty}  - \sigma_{ab}|_{u= -\infty}\big)$ is  non-zero and given by $[\sigma] = \TF (D_{a}D_{b} f)$  (where $\TF$ stands for `Trace Free part of').  This is the \emph{gravitational memory effect}. $[\sigma]$ encodes both the `linear and non-linear'  or  `ordinary and null' memory \cite{memory1,memory2,memory3}.  The effect disappears for connections $\{D\}$ that  tend to the \emph{same} classical vacuum $\{\Do\}$ at both ends of $\scri$. However, in this case the news tensor $N_{ab}$ of $\{D\}$ is severely constrained: it has to satisfy $ [\sigma] \equiv \int\! {\rm d}u\, N_{ab} (u,\theta,\phi) =0$ along \emph{each integral curve of} $n^{a}$, separately, or equivalently, the soft charge $Q_{f}({\rm soft})$ of Eq. (\ref{6}) vanishes identically for all $f$! Therefore, restriction to such connections is physically unreasonable in classical general relativity. There is no reason to expect that this condition would be satisfied at $\scrip$ by the gravitational waves emitted, for example, in a binary coalescence. But, as we will now show, in quantum theory Fock states are subject to this strong restriction.

Since the radiative modes of the full, non-linear theory have been captured in the connections $\{D\}$ without having to make a perturbative expansion, the $\scri$ framework is well suited for asymptotic quantization of the \emph{full, non-linear theory}. Detailed considerations \cite{aa-asym,aa-bib,aa-yau} involving the affine space structure of $\Gammabf$ lead one to introduce smeared News operators $\h{N}(\g)$: 
\be \label{7} \h{N}(\g) :=  -\f{1}{8\pi G}\, \int_{\scri} \rmd^{3}\scri\, \h{N}_{ab} \g_{cd} q^{ac} q^{bd}\,  \ee
where the smearing fields $\g_{ab}$ --like the News tensor-- are transverse, traceless and belong to the Schwarz space of rapidly decaying functions.
(In the final picture $\g_{ab}$ will turn out to be related to shear.) These $\h{N}(\g)$ satisfy the commutation relations
\ba  \label{8} [\h{N}(\g), \, \h{N}(\g^{\prime})] &=& i\hbar\,\, \Omega(\g,\,\g^{\prime}) \, \h{I},\quad \hbox{\rm or, equivalently}\quad\\ 
\hskip-0.0cm  [\h{N}_{ab}(u,\theta,\varphi), \, \h{N}_{a^{\prime}b^{\prime}}(u^{\prime}, \theta^{\prime}, \varphi^{\prime})] &=& 16\pi i G\hbar\,\,\, \delta^{2}(\mathbb{S}^{2}) \,\partial_{u} \delta(u,\,u^{\prime})\,\, \times \nonumber\\
&-&\big( q_{a^{\prime} (a}\, q_{b)b^{\prime}}\, -\,\textstyle{\f{1}{2}} q_{ab}\, q_{a^{\prime}b^{\prime}}\big) \, \h{I}, \label{9}  \ea
%
%
%
%
where $\delta^{2}(\mathbb{S}^{2})$ is the Dirac $\delta$-distribution on the unit 2-sphere. As usual, the algebra  $\A$ generated by these `basic' operators uses only the structure available on the phase space $\Gammabf$.  However, to find the (Fock) representation of $\A$, we need a new ingredient: 
decomposition of $\g_{ab}$ into positive and negative frequency parts. In a Bondi conformal frame, one sets
\be\label{10}  \g^+_{ab}(u,\theta,\phi) = \int_0^\infty\!\!\!\!\rmd \omega\,\, \tilde{\g}_{ab} (\omega,
\theta,\phi)\, e^{-i\omega u}\,  \quad {\rm and}\quad \g^-_{ab} = (\g^+_{ab})^\star\, , \ee
where $\tilde{\g}_{ab} $ is the Fourier transform of $\g_{ab}$ in $u$. Then the News operators can be decomposed into creation and annihilation parts, $\hbar\, \h{a}(\g) = \h{N}(\g^{+})$ and  $\hbar \,\h{a}^{\dag}(\g) = \h{N}(\g^{-})$, so that $\h{N}(\g) = \hbar\,\big(\h{a}(\g) +\h{a}^{\dag}(\g) \big)$.  The vacuum state $|0\rangle$ can then be defined by $\h{a}(\tau) |0\rangle =0$ for all $\tau$ and we will denote the one particle excitations $\h{a}^{\dag}(\g) |0\rangle$ by $|\g\rangle$ or $\g_{ab}$. Commutation relations {(\ref{8})}  imply that the norm of this state is given by: 
\be \label{11} \langle \g | \g\rangle =  \f{i}{\hbar} \,\Omega(\g^{+}, \g^{-})  \equiv \f{1}{8\pi G\hbar}
\oint {\rm d}^{2}S \int_{0}^{\infty} \!\!\!\!{\rm d}\omega \, \omega\, |\t{\g}_{ab} (\omega,\theta,\varphi)|^{2}
.\ee
The `1-particle' Hilbert space $\H$ is the Cauchy completion of the Schwarz space and includes of all $\g_{ab}$ with finite norm (\ref{11}).  The full Fock space $\F$ is generated, as usual, by repeatedly operating on the vacuum by various creation operators. 

One can show that the natural action on $\H$ of the BMS group $\B$  --and hence, of any poincar\'e subgroup $\P$ thereof-- is unitary. Now, irreducible unitary representations of the Poincar\'e group are labelled by mass and spin. In our case, the representation is reducible: it can be decomposed into two irreducible parts. The `right handed sector'  is the subspace\, $\H_{\rm R}$\, of $\H$  consisting of $\g_{ab}$ satisfying $\epsilon^{abc} \ell_{c} q_{bm} \g^{+}_{cn} = i \g^{+}_{mn}$ and the `left handed sector' \,$\H_{\rm L}$\, consists of $\g_{ab}$ satisfying $\epsilon^{abc} \ell_{c} q_{bm} \g^{+}_{cn} = - i \g^{+}_{mn}$, so that $\H = \H_{\rm L} \oplus \H_{\rm R}$. For any Poincar\'e  sub-group $\P$ of $\B$, the mass Casimir has eigenvalue zero on both sectors.  The second Casimir is then the helicity operator: It has eigenvalue +2 on $\H_{R}$ and -2 on $\H_{L}$. In this precise sense we can indeed interpret the excitations created by $\h{a}^{\dag}(\g)$ as gravitons, \emph{even though we are working with full non-linear general relativity}, without any reference to perturbation theory on Minkowski space \cite{aa-asym,aa-bib}.

Recall that in quantum field theory in curved space-times the 1-particle Hilbert space is obtained by an appropriate Cauchy completion of the classical phase space. Therefore, it is useful to further explore the relation between classical states --points $\{D\}$ of $\Gammabf$-- and  1-graviton quantum states $|\g\rangle \equiv \g_{ab}$ in $\H$.  Let us fix \emph{any} classical vacuum $\{ \mathring{{D}} \}$ as the origin in $\Gammabf$. 
Then other points $\{D\}$ are naturally labelled by transverse traceless tensors $\sigma_{ab}$ (which can be interpreted as the shear assigned by $\{D\}$ to the cross sections that are shear-free with respect to $\{ \mathring{{D}} \}$).  One can verify that the News tensor defined by $\{D\}$ is given just by $N_{ab} = 2\mathcal{L}_{n} \sigma_{ab}$. The classical state $\{D\}\in \Gammabf$ defines the 1-particle state $|\sigma\rangle = \h{a}^{\dag}(\sigma)|0\rangle$ (obtained by setting $\g_{ab} = \sigma_{ab}$), {provided} the 1-particle norm (\ref{11}) of $\sigma_{ab}$ is finite.  Thus, $\H$ can now be identified as a sub-space $\Gammabf_{\{\Do\}}$ of $\Gammabf$ consisting of those $\{D\}$ for which $\langle \sigma|\sigma\rangle < {\infty}$. \emph{This is a new,  quintessentially quantum condition that has no classical analog.} In terms of the News tensor, we have:  $\{D\} \in \Gammabf_{\{\Do\}}$  if and only
\be \label{12}  \int_{0}^{\infty} \! \f{\rm{d} \omega}{\omega} \,\big| \t{N}_{ab}|^{2} (\omega,\theta,\varphi) < \infty  \quad \hbox{\rm for each $\theta,\varphi$} \ee
where, as before, the tilde denotes the  Fourier transform. Now, since $\{D\} \in \Gammabf$, $N_{ab}$ is square-integrable in $u$, whence $\t{N}_{ab}$ is square-integrable in $\omega$ and convergence in the ultraviolet is assured. However,  the integral (\ref{12})  is infrared divergent unless $\t{N}_{ab}|_{\omega=0} =0$, or, equivalently, $ \int {\rm d}u \,N_{ab}(u,\theta,\varphi) \equiv 2[\sigma]\, =0$ for all $\theta,\varphi$\,  \cite{aa-asym,aa-bib}.  Recall that \emph{this is a severe restriction from the classical perspective}. In particular,  it implies that  \emph{all} soft charges $Q_{f}({\rm soft})$ of Eq. (\ref{6}) vanish for  every $\{D\}$ that defines a 1-graviton state in $\H$. 

Finally, in these considerations, initial choice of the classical vacuum $\{\Do\}$ is irrelevant: the 1-graviton state $|\g\rangle$ is associated with the entire class of connections $\{D\}$ 
for which $\{D\} -\{\Do\} = \g_{ab}$ for \emph{some} classical vacuum $\{\Do\}$. Put differently,
$|\g\rangle$ corresponds to the whole family of connections $\{D\}$ whose news tensor is given by $N_{ab} = 2\mathcal{L}_{n} \g_{ab}$. \emph{What matters is the news tensor rather than the shear}  because we are working with the Fock representation of News algebra $\A$. Thus, the correspondence between quantum and classical states can be described more succinctly as follows. Consider the subspace $\Gammabf_{0}$ of the full phase space $\Gammabf$ consisting of connections $\{D\}$ for which $\int \rmd u\, N_{ab} = 0$\, or, equivalently, $Q_{f}({\rm soft}) = \int \rmd^{3}\scri \, N_{ab} (D_{c}D_{d}f) q^{ac} q^{bd}  =0$ for all $f(\theta,\varphi)$.
Then the 1-graviton Hilbert space $\H$ is isomorphic with the (Cauchy completion of the) quotient $\Gammabf_{0}/\V$. The Fock vacuum $|0\rangle$ is the coherent state in $\F$ peaked at the origin in $\Gammabf_{0}/\V$ -- i.e., on the entire space $\V$ of classical vacua.

Let us summarize. It is striking that one can construct a Fock representation starting from radiative modes of  \emph{full non-linear general relativity} and, furthermore, using any Poincar\'e subgroup of the BMS group, show that the one particle excitations have zero mass and helicity $\pm 2$. In this precise sense they represent gravitons. However, there is a complementary surprise as well: these \emph{Fock quantum states are severely restricted} from a classical perspective because they refer only to the subspace $\Gammabf_{0}$ on which all soft charges $Q_{f}({\rm soft})$ vanish. One would expect radiation emitted by physically reasonable sources in general relativity to admit a quantum description --say, as a coherent state whose expectation value is peaked at the classical radiation field with small uncertainties. However, this is not generally the case even if the classical radiation field is `tame' --falling off rapidly as $u\to \pm \infty$-- if any of the soft charges $Q_{f}({\rm soft})$  fails to vanish!

\emph{Remark:} Invariant structures underlying the Fock representation can be summarized as follows. The mathematical object supplied by the positive and negative frequency decomposition is a complex structure $J$ on $\Gammabf_{\{\Do\}}$:\, $J\, \g_{ab} := i \g^{+}_{ab} - i \g^{-}_{ab}$. It is compatible with the symplectic structure (\ref{4}) in the sense that $J, \Omega$ provide a  {K\"ahler} structure, which in turn defines the 1-particle Hilbert space $\H$. For details and for a succinct construction of the Fock representation using Weyl operators $\h{W}(\g) = \exp i \h{N}(\g)$, see e.g. \cite{aa-yau}.
 \section{Perturbative S-matrix}
  \label{s5}
Asymptotic quantization of section \ref{s4} provides the appropriate kinematical setup for the S-matrix theory.  In recent years, {there} have been striking advances in this area from different directions; see, e.g., \cite{as-book,as1,ra1,as2,lm,as3,skinner,mcal1,mcal2,as-ir,alas1,ra2,ra3,alas2,alas3} for results in 4-dimensions.  Because of space limitation, will restrict ourselves to topics that the intended audience would be most interested in: role of infrared sectors in the S-matrix theory, proposals of new conservation laws, and (leading-order) Weinberg's soft graviton theorems \cite{sw}, all in perturbative quantum gravity.

Since the S-matrix is a map from states on $\scrim$ to those on $\scrip$, it is natural to ask for physical quantities that are conserved in the process. The issue is subtle because given a physical quantity --such {as} a component of energy momentum-- on $\scrim$, one has to first identify the `same' component at $\scrip$ before asking if they are equal. Let us begin with the classical theory.  Given any cross-section $C$ of $\scri$, one can define the Bondi-Sachs 4-momentum  that associates with any BMS translation $fn^{a} \in \T$ a number $P^{\rm BS}_{f}[C]$, the association being linear in $f$. Thus, the Bondi-Sachs 4-momentum $P^{\rm BS}_{a}[C]$ is a vector in the space dual to $\T$. Given any two cross-sections $C_{1}$ and $C_{2}$ bounding a region $\Delta \scri$, the difference $P^{\rm BS}_{a}[C_{1}] - P^{\rm BS}_{a}[C_{2}]$ equals the flux $F_{a}[\Delta\scri]$ of the Bondi-Sachs 4-momentum carried by gravitational waves across $\Delta\scri$ (obtained by restricting the integral in Eq. (6) to $\Delta\scri$).  This is the Bondi-Sachs balance law. Now, if the space-time is also asymptotically flat at spatial infinity $i^{o}$, one obtains a stronger result  relating $P_{a}^{\rm BS}$ with the ADM 4-momentum $P_{a}^{\rm ADM}: \, P_{a}^{\rm BS} [C] = P_{a}^{\rm ADM} - F_{a}[\Delta_{C}\,\scrip]$, where $F_{a}[\Delta_{C}\,\scrip]$ is the flux of the Bondi-Sachs 4-momentum carried by gravitational waves across the (infinite) region of $\scri$ to the past of the given cross-section $C$ \cite{aa-io}. While the Bondi-Sachs quantities refer to the translation subgroup of $\B$, the ADM 4-momentum refers to the 4-dimensional translation subgroup $\T_{i^{o}}$ in the Spi group $\mathfrak{S}$ at $i^{o}$. Nonetheless it is meaningful to compare these quantities because the structure at $i^{o}$ enables one to set up an isomorphism between $\T\in \B$ and $\T_{i^{o}} \in \mathfrak{S}$.  This result allows for matter sources as well as black holes in the space-time interior and holds for both $\scri^{\pm}$. 

Let us now  restrict ourselves only to \emph{vacuum solutions without black holes}. Then,  
with appropriate fall-off at the two ends of each of $\scri^{\pm}$, the Bondi 4-momentum at $i^{\pm}$ vanishes. In this case, if we move the cross-section $C$ on $\scrip$ to $i^{+}$ and that on $\scrim$ to $i^{-}$, the result implies that the $P^{\rm ADM}_{a}$ equals the total flux $F_{a}[\scri]$ for both $\scri^{\pm}$. In particular then, the classical S-matrix satisfies a conservation law: Energy momentum $F_{a}[\scrim]$ carried by the incoming gravitational waves across $\scrim$ equals the flux $F_{a}[\scrip]$  carried across $\scrip$. The question is whether a similar law holds also for the incoming and outgoing Bondi-Sachs supermomentum. The question is meaningful because the structure at $i^{o}$ again allows us to set a 1-1 correspondence between supertranslations on $\scri^{+}$ with those at $\scri^{-}$.%
 \footnote{This result  first uses the fact that the conformally completed metric at $i^{o}$ is $C^{0}$ to set an isomorphism between the generators of $\scri^{\pm}$ and then uses the fact that the ADM 4-momentum is time-like to set up a correspondence between Bondi conformal frames at $\scri^{\pm}$. The second step is necessary because supertranslations are characterized by \emph{conformally weighted} functions.} 
Recently, results from perturbative quantum gravity have been used to argue that  there is indeed such a conservation law.
 
\subsection{Beyond the Fock representation: Infrared sectors I}
  \label{s5.1} 

As we discussed at the end of section \ref{s4}, the 1-graviton Hilbert space $\H$ arises only from those classical radiative modes $\{D\}$ for which all soft charges $Q_{f}({\rm soft})$ of (\ref{6})  vanish. Therefore, one  might intuitively expect  the matrix elements of soft charge operators $\h{Q}_{f}({\rm soft})$ to vanish on the Fock space $\F$. We will now show that this is the case. Note first that,  since $\h{Q}_{f}({\rm soft}) = \int  \rmd^{3}\scri \,\h{N}_{ab} (D_{a} D_{b} f)q^{ac} q^{bd}$, and $\TF(D_{a}D_{b}f)$ has zero frequency, we need a prescription to define the action of this operator on $\F$. A natural avenue is to look for the finite canonical transformation generated by the classical soft charge $Q_{f}(\rm{soft})$ on $\Gammabf$ and define $\exp i[\h{Q}_{f}({\rm soft})]$ as the unitary operator that lifts this action to $\F$. The canonical transformation is $\{D\} \to \{D\} + \TF(D_{a}D_{b}f)$. Since the subspaces $\Gammabf_{0}$ and $\V$ of $\Gammabf$ are invariant under this map, the action naturally descends to the quotient $\Gammabf_{0}/\V$. It is easy verify that the action on the quotient is just the identity. Since $\Gammabf_{0}/\V$ is isomorphic with the 1-graviton Hilbert space $\H$, the map  lifts to quantum theory  and is just the identity operator. In this precise sense, \emph{in the Fock representation of the News algebra, the entire $\F$ is an eigenspace of $\h{Q}_{f}({\rm soft})$ with zero eigenvalue.}

New representations of the News algebra $\A$ \emph{with non-zero soft charges} arise as follows. Fix a generic radiative mode $\{\b{D}\}$  in the \emph{full} phase space $\Gammabf$ and define an automorphism (i.e. structure preserving map) on the quantum algebra $\A$ induced by the map
\be  \label{14} \h{N}(\g) \to \Lambda_{\{\b{D}\}} \h{N}(\g) := \h{N}(\g) + \big({\int}_{\!\!\scri}{\rm d}^{3}\scri\,\b{N}_{ab} \,\g_{cd} q^{ac}q^{bd} \, \big) \,\,\h{I} \ee
where $\b{N}_{ab}$ is the news tensor of the given $\{\b{D}\}$. This automorphism is {unitarily} implementable if and only if $[\b\sigma]=0$,\, i.e. $Q_{f}({\rm soft})|_{\{\b{D}\}} =0$ for all $f$. In this case, the unitary map $U$ satisfies $U|0\rangle = \exp i \h{N}(\g) |0\rangle = |C_{\{\b{D}\}}\rangle$, the coherent state in $\F$ peaked at  $\{\b{D}\}$, and for any operator $\h{A}\in \A$, 
\be \label{15} \langle 0| (\Lambda_{\{\b{D}\}}  \h{A}) |0\rangle = \langle 0| U^{\dag} \h{A} U |0\rangle = \langle C_{\{\b{D}\}} | \h{A} |C_{ \{\b{D}\} }\rangle. \ee

Now suppose $[\b\sigma] \not=0$. Then the Fock norm of the `would be' coherent state is infinite, whence $\Lambda_{\{\b{D}\}}$ is not unitarily implementable in the Fock space. Instead, one can use the vacuum expectation value (or the positive linear function)  ${}_{\{\b{D}\}}\langle 0|\h{A} |0\rangle_{\{\b{D}\}} := \langle 0|(\Lambda_{\{\b{D}\}}  \h{A}) |0\rangle$ to construct a \emph{new} representation of $\A$ (via Gel'fand-Naimark-Segal construction \cite{gns}).
This representation is unitarily \emph{inequivalent} to the Fock representation we began with: It is  `displaced' relative to the Fock representation in that,  the Fock vacuum $|0\rangle$  is replaced by a `coherent state' peaked on $\{ \b{D} \} \not\in \Gammabf_{\{\Do\}}$ (see (\ref{12})). 
The $n$-particle states of the new Hilbert space $\F_{[\b{\sigma}]}$  are excitations over this `displaced' vacuum, obtained by the action of the creation operators \cite{aa-asym,aa-bib}. Let us compute the matrix elements of the soft charge $\h{Q}_{f}({\rm soft})$ in this new representation: They are given by the matrix elements of $\Lambda_{\{\b{D}\}}  [\h{Q}_{f}({\rm soft})] =   [\h{Q}_{f}({\rm soft})] + q_{f}\,\h{I}$ between Fock states,  where $q_{f} = \int \rmd^{3} \scri \,\b{N}_{ab} \TF(D_{c}D_{d} f)\, q^{ac} q^{bd}$, the value of the classical soft charge of $\{\b{D}\}$.  Thus the displaced Fock space $\F_{[\b{\sigma}]}$ is the eigenspace of soft charge operators $\h{Q}_{f} ({\rm soft})$ with eigenvalues $q_{f}$ \emph{for all $f$}. Consequently we can label  $\F_{[\b{\sigma}]}$ also as $\F_{\{q(f)\}}$. Each of these displaced Fock spaces $\F_{\{q(f)\}}$ is `as large' as the Fock space $\F$ we began with. However, states in $\F_{\{q(f)\}}$ carry an additional label $[\b\sigma]$, or, equivalently $\{q_{f}\}$.  The label is unnecessary for the standard Fock states since $[\b\sigma]=0=\{q_{f}\} $ for them.

What role do these new $\F_{\{q(f)\}}$ have in the S-matrix theory? It had been argued sometime ago  \cite{aa-asym,aa-bib} that elements of  $\F_{\{q(f)\}}$ are the analogs of the dressed asymptotic states that are necessary in QED  \cite{fk} to go beyond finiteness of cross-sections, and make the S-matrix itself  infrared finite. The recently proposed conservation laws \cite{as2,as1} have now been used to capture this relation in a concrete manner \cite{as-ir,ra2,ra3}. Because of space limitation we will have to gloss over some technical subtleties. In particular, we will present the overall picture using just the gravitons (thus overlooking the fact that some of the results have been obtained only when one uses massive particles in place of `hard' graviton states). 

Recall that on the full phase space $\Gammabf$, the Hamiltonian $H_{f} = Q_{f}({\rm hard})+ Q_{f}({\rm soft})$ (given in Eq (5)) generates the supertranslation $fn^{a}$  and represents the total supermomentum.  $Q_{f}({\rm hard})$  --and hence $H_{f}$-- can also be readily promoted to quantum operators on all  $\F_{\{q(f)\}}$. Let us consider the enlarged space of asymptotic states on $\scri^{\pm}$ on which the perturbative S-matrix is well-defined in the infrared \cite{ra1}, \emph{assume} that  supermomenta are conserved under S --i.e., $ [\h{\H}_{f},\, S] =  [\h{Q}_{f}({\rm soft}) + \h{Q}_{f}({\rm hard}),\, S] =0$-- and work out the implications.  Suppose we have a $m$ particle incoming state $|\inn \rangle\equiv |\vp_{1} \ldots \vp_{m};\, \{q^{\inn}(f)\} \rangle \in \F_{\{q^{\inn}(f)\}}$ on $\scrim$, and an $n$ particle outgoing state $|\out\rangle \equiv |\vk_{1} \ldots \vk_{n};\, \{q^{\out} (f)\}\rangle \in 
\F_{ \{q^{\out}(f)\}} $ on $\scrip$. Then, taking matrix elements of the assumed dynamical conservation law in these states, we have
\ba \label{16} &&\hskip1.5cm \langle \out| \,[\h{Q}_{f}({\rm soft}) ,S]\, |\inn\rangle\,\, +\,\,\langle \out\, |[\h{Q}_{f}({\rm hard}) ,S]\,| \inn\rangle =0. \\
&& \hskip-2.8cm{\hbox{\rm Definitions of the charge operators then imply  \cite{ra2}:}}\nonumber\\
&&\hskip-0.9cm \label{17} \Big[\big(q_{f}^{\out}({\rm soft}) - q_{f}^{\inn}({\rm soft})\big)\, - \big(\sum_{i=1}^{m}  |\vp_{i}|\, f(\h{\vp}_{i})  - \sum_{j=1}^{n}  |\vk_{j}|\, f(\h{\vk}_{j}) \big) \Big]\,\, \langle \out| S| \inn\rangle  =0,  \ea
where on the right side of (\ref{17}) the function $f$ is evaluated at the point $\theta,\varphi$ on the 2-sphere determined by unit 3-vectors $\h{\vp}_{i} = \vp_{i}/|\vp_{i}|$ and $\h{\vk}_{j} = \vk_{j}/|\vk_{j}|$.  As we discuss below, this is a strong constraint on the S-matrix, implied by our assumption that all matrix elements of the commutator $[\h{\H}_{f},\, S]$ vanish. Now, calculations of S-matrix elements have shown that Eq. (\ref{17}) does hold order by order  in perturbative gravity (provided one includes all loop diagrams; it does not hold at the tree level). Therefore one can reverse the argument and conclude that the matrix elements of the commutator $ [\h{\H}_{f},\, S] $ vanish. In this sense, then, there is a conservation law for supermomentum flux in perturbative quantum gravity.

Note that Eq (\ref{17}) implies that the S-matrix elements between the in-states $|\inn\rangle$ and the out-states $|\out\rangle $ vanishes unless the momenta  $\vk_{j}$ and the $\vp_{i}$ of `hard' particles in the two states are related to the soft charges \emph{such that the square bracket on the left side vanishes}.  This is a striking constraint. In particular, if the `in' and `out' states both belong to the Fock representation, then $q^{\inn}_{f}({\rm soft}) =q^{\out}_{f}({\rm soft}) =0$. So the full transition amplitude $\langle \out| S| \inn\rangle$ (including loop diagrams) is \emph{identically zero} unless the momenta are finely tuned so that the quantity in the second round bracket in (\ref{17}) vanishes \emph{for all} functions $f(\theta,\phi)$. This is an extremely strong restriction (satisfied if the `in' and `out' states correspond to trivial scattering).  We had seen from classical considerations that Fock states are too  restrictive because that theory does not incorporate radiation fields produced by realistic sources. We now see that the restriction to Fock sectors is too restrictive also from perturbative quantum S-matrix considerations.\medskip

\emph{Remark:} Because the action of the BMS group $\B$ on $\Gammabf$ leaves the subspaces $\Gammabf_{0}$  and $\V$ invariant, it can descend to the 1-graviton Hilbert space $\H$ in the Fock representation. We used this fact to assign mass and helicities to $\H$. However, the displaced subspaces $\Gammabf_{[\b\sigma]}$ of $\Gammabf$ consisting of $\{\b{D}\}$ with $\b\sigma_{ab}|_{i^{+}} - \b\sigma_{ab}|_{i^{o}} = [\b\sigma] \not= 0$ are not left invariant by \emph{any} Poincar\'e subgroup $\P$ of $\B$ on $\scrip$; they are left invariant only by the translation subgroup $\T$ of $\P$. Therefore Lorentz subgroups $\L$  do not have a well-defined action on the displaced representations of $\A$ and the notion of angular momentum loses its clear-cut meaning.  A second difference from the Fock representation is that just as $\Gammabf_{[\b\sigma]}$ does not admit any $\{D\}$ with zero energy, $\ul{\F}_{\{q(f)\}}$ does not admit any state $|\psi\rangle$ with zero energy (although it admits states with arbitrarily small energy). These differences capture the `price' one has to pay to make the quantum S-matrix well-defined by `dressing' the asymptotic states.

 \subsection{Beyond the Fock representation: Infrared sectors II}
 \label{s5.2}

Our discussion of infrared sectors of section \ref{s5.1} and the subsequent results (\ref{16}) - (\ref{17}) did not make  a direct reference to Weinberg's soft theorem in perturbative quantum gravity \cite{sw}. We will now introduce a \emph{different family} of infrared sectors to summarize our understanding of the relation between the soft theorem and supertranslations \cite{as1,as-book}. The fact that distinct infrared sectors exist and are necessary for these two different considerations does not appear to be appreciated in the literature.
 
As we saw in section \ref{s4}, the Fock representation of the News algebra $\A$ admits a unique vacuum $|0\rangle$; the classical vacuum degeneracy disappears because the representation is sensitive to the `News content' of connections $\{D\}$ rather than their `shear content'. Now, in perturbative quantum gravity, the background Minkowski metric provides a preferred classical vacuum $\{\Do\}_{0}$. We will now show that  the `shear content' \emph{with respect to this} $\{\Do\}_{0}$ can be captured in quantum theory by replacing $\A$ with the algebra $\ul\A$ generated by  shear operators, $\h{\sigma}(u,\theta,\varphi)$ that  satisfy 
\be  [\h{\sigma}(u,\theta,\varphi), \, \h{\sigma}(u^{\prime}, \theta^{\prime}, \varphi^{\prime})] = 4\pi i G\, \hbar\, \delta^{2}(\mathbb{S}^{2}) \, \Delta(u,\,u^{\prime}) \big( q_{a^{\prime} (a} q_{b)b^{\prime}} - \textstyle{\f{1}{2}} q_{ab}\, q_{a^{\prime}b^{\prime}}\big) \, \h{I}, \label{18}\ee
where $\Delta$ is the step function. (As in the classical theory, $\h{N}_{ab} = 2 \mathcal{L}_{n} \h\sigma_{ab}$; see (\ref{9}).) The Fock representation of $\ul\A$ can be obtained by writing $\h\sigma_{ab}$ as a sum of creation and annihilation operators using the positive and negative frequency decomposition on $\scri$.  As a result,  in addition to $\h{N}_{ab}$, now we have access also to the shear operators $\h{\sigma}_{ab}$.  Note the differences from the Fock representation on $\A$: The vacuum state $|0\rangle$ is now peaked at a specific classical vacuum $\{\Do\}_{0}$  --rather than on the space $\V$ of all classical vacua-- and the 1-graviton Hilbert space $\Hs$ can be identified with the space $\Gammabf_{\{\Do\}_{0}}$ of all connections $\{\ub{D}\} $ that tend to $\{\Do\}_{0}$ as $u \to \pm \infty$  --rather than with $\Gammabf_{0}/ \V$. 

Let us now consider any connection $\{\ub{D}\}\in \Gammabf_{\{\Do\}_{0}}$ which can be labelled by the shear tensor $\ul\sigma_{ab}$ with respect to $\{\Do\}_{0}$. Then $|C_{\{\ub{D}\}} \rangle := \exp i[{\h{N}(\ul\sigma)}]\,|0\rangle$ is a coherent state in the Fock space $\ul\F$ of $\ul\A$, peaked at the connection $\{\ub{D}\}$, satisfying $\langle C_{\{\ub{D}\}} | \h\sigma_{ab}  |C_{\{\ub{D}\}} \rangle = \ul{\sigma}_{ab}$. Next, consider a 1-parameter family of connections 
$\{\ub{D}\}(\lambda)$  in $\Gammabf_{\{\Do\}_{0}}$, labelled by $\ul{\sigma}_{ab}(\lambda)$, satisfying $\lim_{\lambda\to \infty} \ul\sigma_{ab} (\lambda) = \TF(D_{a}D_{b} \ub{f})$,  where $\ub{f}$ is a supertranslation.  (An example is provided by $\ul{\sigma}_{ab} (\lambda) = [\exp -(u^{2}/\lambda^{2}) ]\, {\TF}(D_{a} D_{b}\,\ub{f})$.) Let us denote by $\{{\Do}\}_{\ub{f}}$ the image of $\{\Do\}_{0}$ under this supertranslation. Then, intuitively, as $\lambda\to \infty$, the limiting state $\lim\, [\exp i[{\h{N}(\ul\sigma)(\lambda)}]\,|0\rangle]$ would be a coherent state peaked at $\{{\Do}\}_{\ub{f}}$. It is not in the Fock space $\ul\F$.\,  Rather, it represents  a \emph{`condensate of gravitons with zero frequency/energy'} in which the expectation value of $\h{\sigma}_{ab}$ is given by $\lim \ul\sigma_{ab}(\lambda) = \TF(D_{a} D_{b}\,\ub{f}\,)$.  We will denote this state by $|0\rangle_{\ub{f}}$. By acting repeatedly by creation operators on  $|0\rangle_{\ub{f}}$, one generates a Hilbert space $\F_{\ub{f}}$ carrying a \emph{new} representation of $\ul\A$ that is unitarily inequivalent to the Fock representation.%
\footnote{These considerations can be made precise by considering a 1-parameter family of automorphisms $\h\sigma(u,\theta,\varphi) \to \h\sigma(u,\theta,\varphi) + \ul\sigma(u,\theta,\phi) \h{I}$ on $\ul\A$, following the procedure used in section \ref{s5.1}. Note also that, since $\TF(D_{a}D_{b}f)=0$ for translations, the representations are really labelled by the `pure supertranslation part' of $f$.}\smallskip

What is the soft charge associated with these new representations? Consider first  the soft charge $Q_{\ub{f}}({\rm soft})$ on the classical phase space $\Gammabf$. Recall that the finite canonical transformation it generates  is $\{D\} \to \{D\} + \TF(D_{a} D_{b} \, \ub{f})$.  We found in section \ref{s5.1} that the corresponding unitary map $\exp i(\h{Q}_{\ub{f}} ({\rm soft}))$ is the identity operator on the Fock representation of the News algebra $\A$  because the News tensor of $\{D\}$ is the same as that of its image. On the other hand, shear (with respect to $\{\Do\}_{0}$) of a connection $\{D\}$ is \emph{not} the same as that of its image: we have $\sigma_{ab} \to \sigma_{ab} + \TF(D_{a} D_{b}\, \ub{f})$.  Therefore the lift  of this canonical transformation to the quantum theory of \emph{shear operators} now under consideration is non-trivial. The lift,  $\exp i(\h{\ub{Q}}_{\ub{f}}({\rm soft}))$, maps each representation of the shear algebra $\ul\A$ to another, `shifted' representation. Since $\ul{\sigma}(\lambda) \to \TF(D_{a}D_{b}\, \ub{f})$  as $\lambda\to \infty$, the action of $\exp i(\h{\ub{Q}}_{\ub{f}}({\rm soft}))$ can be obtained using the one parameter family $\ul{\sigma}_{ab} (\lambda) $ considered above: For example, $ \exp i(\h{\ub{Q}}_{\ub{f}}({\rm soft})) |\vk_{1} \ldots \vk_{n} \rangle = \lim \big(\exp i[\h{N}(\ul{\sigma})(\lambda)]\, |\vk_{1} \ldots \vk_{n} \rangle\big) =: |\vk_{1} \ldots \vk_{n} \rangle_{\ub{f}} \in \F_{\ub{f}}$. In particular, the Fock vacuum $|0\rangle \in \F$ is mapped to $|0\rangle_{\ub{f}} $. 
\medskip

We can now explore consequences of conservation of supermomentum. Let us restrict ourselves to the \emph{tree-level} S-matrix  elements ${}_{0}\langle \out | \ub{S} |\inn\rangle_{0}$, 
which are  well-defined for the incoming  and outgoing graviton states  $|\inn\rangle_{0} = | \vp_{1} \ldots \vp_{m}\rangle_{0}$ and $|\out\rangle_{0} =  | k_{1},\ldots \vk_{n} \rangle_{0}$ in the Fock spaces $\F_{0}$ on $\scri^{\mp}$. Suppose the supermomenta are conserved in this scattering process \emph{in the sense that} 
\ba &&\label{19} \hskip1cm  {}_{0}\langle \out| \,\Big[\big(\exp i[\h{\ub{Q}}_{\ub{f}}({\rm soft})+ \h{{Q}}_{\ub{f}}({\rm hard})] \big) ,\, \ub{S} \Big]\, |\inn\rangle_{0}\, =0\\
&& \hskip -3.4cm \hbox{\rm Since the two charge operators commute, we have:}\nonumber\\
\label{20} && \hskip1cm {}_{0}\langle \out| \, \big(\exp -i\h{\ub{Q}}_{\ub{f}}({\rm soft})\big)\, \ub{S}\,  \big(\exp i\h{\ub{Q}}_{\ub{f}}({\rm soft})\big) \,| \inn\rangle_{0}\, =\, \nonumber\\
&&\hskip1cm \,  {}_{0}\langle\out|\, \big(\exp i\h{{Q}}_{\ub{f}}({\rm hard})\big)\, \ub{S}\,  \big(\exp -i\h{{Q}}_{\ub{f}}({\rm hard})\big) \,|\inn\rangle_{0}\, .\\
&&\hskip-3.4cm {\hbox{\rm Definitions of the two charge operators now imply}} \nonumber\\
&&\hskip-0.9cm \label{21} {}_{\ub{f}}\langle {\out}|\ub{S}|{\inn}\rangle_{\ub{f}} = \Big(\exp i \big[\sum_{i=1}^{m}
|\vp_{i}|\, f(\h{\vp}_{i})  - \sum_{j=1}^{n}  |\vk_{j}|\, \ub{f}(\h{\vk}_{j}) \big]\,\Big) \,  {}_{0}\langle\out|\ub{S}|\inn\rangle_{0}\, . \ea
 Left side of (\ref{21}) is the scattering amplitudes for Fock states of (hard) gravitons in presence of a condensate of soft (i.e. zero energy) gravitons.  It is  related to the scattering amplitude between the same (hard) gravitons in absence of soft gravitons via a simple phase factor that refers only to the momenta of the incoming and outgoing particles, and the function $f$ that characterizes the soft condensate. Eq. (\ref{21}) is a version of the tree level Weinberg's \cite{sw} soft theorem for gravitons  (to leading order).  Thus, conservation of supermomentum  in the sense of (\ref{19}) implies the soft theorem (\ref{21}). Since we know  from tree level perturbative calculations that (\ref{21}) holds,  we can reverse the argument to arrive at  (\ref{19}). In this sense there is equivalence between the leading order soft theorem and the Ward identity associated with BMS supertranslations. The central idea behind this argument appeared in \cite{as1,as2,as3}; our discussion sharpens the argument by bringing out subtleties that lead to two different  soft charge operators $\h{Q}_{f}$ and $\h{\ub{Q}}_{\ub{f}}$ and existence of two distinct sets of infrared sectors.\medskip

\emph{Remarks:}\\
\noindent 1. Recall that the soft charge operator is defined through a limiting procedure, $\exp i(\h{\ub{Q}}_{\ub{f}}({\rm soft})) |\psi\rangle = \lim \big(\exp i[\h{N}(\ul{\sigma})(\lambda)]\, |\psi\rangle\big)$ where $\exp i[\h{N}(\ul{\sigma})(\lambda)]$ is a well-defined operator on the Fock space $\F_{0}$.  Thus, in the left hand sides of  (\ref{19}) and (\ref{20}), one replaces $\exp i(\h{\ub{Q}}_{\ub{f}} ({\rm soft}))$ by $\exp i[\h{N}(\ul{\sigma})(\lambda)]$,  calculates the matrix element and then takes the limit to arrive at (\ref{21}). \medskip

\noindent 2. At first it may seem puzzling that there are two soft charge operators. In the classical theory we have a single soft charge observable $Q_{f}({\rm soft})$ given by (\ref{6}). On the entire phase space $\Gammabf$ it generates the (finite) canonical transformation $\{D\} \to \{D\}_{f} = \{D\} + \TF (D_{a}D_{a}f)$.  In section \ref{s5.1} we found that the lift  $\exp i[\h{Q}_{f}({\rm soft})]$ of this action to the Fock space $\F$ of the News algebra $\A$ is identity because $\{D\}$ and  $\{D\}_{f}$ have the same news. The vacuum state $|0\rangle\in \F$, for example, is the coherent state corresponding to the \emph{entire subspace} $\V$ of classical vacua (all of which have zero News).  Therefore, although  an individual classical vacuum  $\{\Do\}$ is mapped to a distinct classical vacuum $\{\Do\}_{f} = \{\Do\} + \TF (D_{a}D_{a}f)$, the quantum vacuum $|0\rangle$ is invariant under this transformation: $\exp i[\h{Q}_{f}({\rm soft})]  |0\rangle = |0\rangle$. In section \ref{s5.2}, by contrast, we lifted the action to the Fock representation of the shear algebra $\ul\A$ which is sensitive to the shear content of connections, not just to the news content. The Fock vacuum $|0\rangle_{0}$  is peaked at the `canonical' classical vacuum $\{\Do\}_{0}$ of perturbative quantum gravity, not on the entire space $\V$ of all classical vacua. Since $\{\Do\}_{0} \to \{\Do\}_{f}$ under the canonical transformation generated by $Q_{\ub{f}}({\rm soft})$, the vacuum $|0\rangle_{0} \in \ul\F$ is sent to $\exp i[\h{Q}_{\ub{f}}({\rm soft})] |0\rangle_{0} = |0\rangle_{\ub{f}} \in \ul\F_{\ub{f}}$, and states $|\Psi\rangle \in \ul\F$ are sent to  $|\Psi\rangle_{\ub{f}} \in \ul\F_{\ub{f}}$. \medskip

\noindent 3. At first there appears to be some tension between Eqs.  (\ref{17}) and (\ref{21}). If we restrict ourselves to in and out state in the standard Fock space, as we emphasized in section \ref{s5.a},  (\ref{17}) implies that, generically,  the matrix element $\langle \out|S|\inn\rangle$ would vanish, while the matrix element ${}_{0}\langle \out| \ub{S} |\inn\rangle_{0}$ of (\ref{17}) does not. This difference arises because ${}_{0}\langle \out| \ub{S} |\inn\rangle_{0}$ are tree-level S-matrix elements, while $\langle \out|S|\inn\rangle$ include virtual loops which suppress the transition amplitude completely.\medskip

\noindent 4. Finally, there is general intuition that the classical scattering map $S_{\rm cl}$ of full general relativity would generically send in-states with $[\sigma]=0$ on $\scrim$ to out states with $[\sigma]\not=0$ on $\scrip$. This intuition is compatible with result (\ref{17})  on S-matrix elements  $\langle \out|S|\inn\rangle$  but at odds with the result (\ref{21})  on the tree-level matrix elements\, ${}_{0}\langle 0|\ub{S}|0\rangle_{0}$. This suggests that the supermomentum conservation law in (\ref{16}) - (\ref{17})  would be of direct relevance to scattering in classical general relativity, in contradistinction with the conservation law in  (\ref{19}) - (\ref{21}) that is more directly related to the soft theorem. \medskip

\noindent 5. While we do have distinct representations of the shear algebra $\ul\A$, they all provide the representations of the News algebra that is unitarily equivalent to the Fock representation in section \ref{s5.1}. Therefore, 
unlike with the infrared sectors of section \ref{s5.1},  the Poincar\'e subgroup $\P$ associated with $\{\Do\}_{\ub{f}}$ has a well-defined action on $\F_{\ub{f}}$ for all $\ub{f}$, and  there is no difficulty with angular momentum.

 \section{Discussion}
\label{s6}

The structure of the gravitational field at null infinity is astonishingly rich both in classical and quantum theory, displaying an elegant and deep interplay between geometry and physics. At the kinematical level, the structure has been well understood for several decades. In particular, in the classical theory the precise relation between BMS supertranslations, classical vacua and gravitational memory had been spelled out and reliable expressions of BMS supermomenta and their fluxes had been obtained by early 1980s. In quantum theory, Fock representation was constructed for full non-linear general relativity, its physical limitation was understood, and infrared sectors (discussed in section \ref{s5.1}) had also been introduced then. However, there were no detailed results on quantum dynamics.

Recent developments have opened avenues to understand \emph{dynamics} via perturbative scattering theory. Progress has occurred along two broad avenues, discussed in sections \ref{s5.1} and \ref{s5.2}: (1) Detailed calculations have been performed to show  that infrared sectors of \ref{s5.1} are necessary to make the perturbative S-matrix elements well-defined in the infrared.  In addition, there are relations between S-matrix elements that are equivalent to conservation of supermomentum fluxes across $\scri^{\pm}$ in perturbative quantum gravity; and, (2) Interplay between (leading order) soft theorems and the Ward identities associated with the supertranslation symmetry have been brought to forefront.  However, as we saw, the infrared sectors and soft charge operators used in these two explorations are conceptually quite different. In particular, Weinberg's soft theorems do not play a direct role in the considerations of section \ref{s5.1}, and the infrared sectors of section \ref{s5.1} play no direct role in the discussion of the soft theorems of section 
\ref{s5.2}. We hope this clarification will help advance the field further.

The possibility of new conservation laws and insights into the infrared behavior of the gravitational field are exciting  because of they bring together very different ideas and results. There is no doubt that the currently intense activity in this area will continue in the coming years.  However, one should also keep in mind that so far arguments are restricted to perturbative scattering  theory on Minkowski space-time. In full general relativity, of necessity, classical scattering theory can be established only for `small' data in a neighborhood of Minkowski space because for more general data black holes can form and $\scri^{\pm}$ cease to provide complete past and future boundaries for the S-matrix theory even for zero rest mass fields. For data considered so far in the global existence theorems, the total flux of `pure' supermomentum \emph{vanishes identically} across both $\scri^{\pm}$. Therefore this analysis \emph{does not} provide non-trivial support for the newly proposed conservation laws. However, other recent results suggest that it may be possible to establish new conservation laws using a Hamiltonian framework \cite{mcre,mhct}  (see also the older work \cite{mhml}). Note also that even apparently tame, low frequency initial data at $\scrim$ can lead to gravitational collapse and black hole formation, if there is appropriate focusing \cite{dc,xa}. Therefore,  the S-matrix between $\scri^{\pm}$ may not capture all the interesting physics even if one stays well away from the Planck regime for incoming states. Progress will only be enhanced if one keeps track of such caveats.

 \section*{Acknowledgments}
This article was motivated by questions raised by several colleagues, especially Eanna Flanagan, John Friedman and Bob Wald during the \emph{GR21 conference}. We have also benefited from discussions with participants of the 2018 \emph{Solvay workshop on Infrared Physics} and extensive discussions with Ashoke Sen, Prahar Mitra and Lionel Mason on soft theorems and related issues. Referee's comments and correspondence with experts on the first version of this paper led to several clarifications in the final version. This work is supported in part by the NSF grants  PHY-1505411, PHY-1806356,  the Urania Stott fund of Pittsburgh foundation, the Eberly research funds of Penn State and PEDECIBA and SNI of Uruguay. 

\end{document}